\documentclass[aps,twocolumn,prc,amsmath,amssymb,showpacs,floatfix]{revtex4}

\usepackage{graphicx}
\usepackage{dcolumn}
\usepackage{bm}

\newcommand{\bei}{\ensuremath{{^8}\mathrm{B}}}
\newcommand{\bese}{\ensuremath{{^7}\mathrm{Be}}}
\newcommand{\uEMD}{\ensuremath{\mathrm{EMD}}}  
\newcommand{\uE}{\ensuremath{\mathrm{E}}}  
\newcommand{\ud}{\mathrm{d}}
\newcommand{\urms}{\ensuremath{\mathrm{rms}}}
\newcommand{\dbde}{\ensuremath{\frac{\mathrm{d} B (\mathrm{E}\lambda)}
{\mathrm{d}E}}}                            

\begin{document}


\title{Radiative capture and electromagnetic dissociation involving
loosely bound nuclei: the $^\mathbf{8}$B example} 


\author{C. Forss\'en}
\email[]{c.forssen@fy.chalmers.se}
\author{N.B. Shul'gina}
\altaffiliation[Permanent address:~]
{The Kurchatov Institute, 123182 Moscow, Russia}
\author{M.V. Zhukov}
\affiliation{Chalmers University of Technology and G\"oteborg
University\\
S--412~96 G\"oteborg, Sweden}


\date{\today}

\begin{abstract}
Electromagnetic processes in loosely bound nuclei are investigated using
an analytical model. In particular, electromagnetic dissociation of
\bei\ is studied and the results of our analytical model are compared to
numerical calculations based on a three-body picture of the \bei\ bound
state. The calculation of energy spectra is shown to be strongly model
dependent. This is demonstrated by investigating the sensitivity to the
rms intercluster distance, the few-body behavior, and the effects of
final state interaction. In contrast, the fraction of the energy
spectrum which can be attributed to E1 transitions is found to be almost
model independent at small relative energies. This finding is of great
importance for astrophysical applications as it provides us with a new
tool to extract the E1 component from measured energy spectra. An
additional, and independent, method is also proposed as it is
demonstrated how two sets of experimental data, obtained with different
beam energy and/or minimum impact parameter, can be used to extract the
E1 component.
\end{abstract}


\maketitle

\section{\label{sec:intro}Introduction}
The properties of loosely bound nuclei have been studied in nuclear
physics for a number of years. In particular some electromagnetic
processes, such as certain charged particle capture reactions, are very
interesting in themselves as they are of vital importance in
astrophysical scenarios. Unfortunately, at stellar energies, the cross
sections for these reactions are very small due to the Coulomb barrier
and direct measurements are therefore extremely difficult. Instead one
has to rely on theoretical extrapolations from experimentally accessible
energies down to stellar ones. An alternative, indirect method to
investigate radiative capture reactions, is to study electromagnetic
dissociation (EMD) on heavy targets~\cite{bau96:46}. This technique
gives an enormous increase in yield due to the huge amount of virtual
photons produced by the high-$Z$ target, the more favorable phase space
factor, and the possibility to use thicker targets. In principle it
should therefore be possible to measure the cross section at very low
relative energies. There are, however, also disadvantages with the
indirect method; the most important being the admixture of
$\gamma$-transitions with different multipolarities whereas the direct
capture process, in most cases, should be completely dominated by E1
transitions. Since the cross sections for radiative capture and
photo-dissociation are related via detailed balance for each separate
multipole, it is necessary to have knowledge of the strengths of
different multipole transitions in the EMD reaction.

The problem of extracting the E1 contribution from a measured EMD energy
spectrum remains a challenge to the nuclear physics community. In
Ref.~\cite{esb95:359} it was proposed to study the angular or momentum
distributions of the breakup fragments. The idea was to employ the fact
that interference terms, between E1 and E2 excitation amplitudes, will
produce an asymmetry in these distributions. This method was, e.g., used
in the analysis of \bei\ EMD~\cite{dav01:63} where the E2 excitation
amplitude, calculated within first-order perturbation theory, was
renormalized in order to reproduce the asymmetry of the measured \bese\
longitudinal momentum distribution. However, as was already noted in
Ref.~\cite{esb95:359}, the asymmetry due to E1-E2 interference strongly
depends on the final state interaction (FSI) between the breakup
fragments; or in other words, on the structure of the continuum up to
relatively large energies. Moreover, terms which contribute to the
asymmetry do not contribute to the integrated cross section from which
the astrophysical $S$-factor is extracted. Finally, for low beam
energies, higher-order dynamical effects will lead to a reduction of the
asymmetry~\cite{esb96:600}. Therefore we conclude that, if we are
interested in astrophysical applications of EMD experiments, it is
desirable to look for more stable, and less model dependent,
characteristics than the asymmetries of angular and momentum
distributions. In this paper we will present two novel methods to
extract the E1/E2 ratio from EMD experiments.

We will use an analytical approach based on a two-cluster picture of the
nucleus, but the effects of many-body structure will also be
included. Our approach will be general in the sense that both
neutron-rich and proton-rich systems can be studied. This model was
first presented in~\cite{for02:549} while similar approaches also exist
for one-neutron~\cite{ber88:480,ots94:49} and
two-neutron~\cite{pus96:22,for02:697,for02:706} halo nuclei. Although
advanced numerical investigations are readily performed utilizing
present day computer power, an analytical approach might have an
advantage when exploring general physics features, and the sensitivity
to different model assumptions.

The physics case which will be investigated throughout this paper is the
\bei\ nucleus. The interest in \bei\ stems from its key role in the
production of high-energy solar neutrinos. It is well-known that the
probability for the reaction $\bese(p,\gamma)\bei$ at solar energies
strongly depends on the structure of \bei\ and, in particular, on the
asymptotics of the valence proton wave function (WF). This reaction has
been studied indirectly through EMD, using a radioactive \bei\ beam
impinging on a heavy
target~\cite{cor02:529,dav01:63,iwa99:83,kik97:391}. Note that the
question of E2 contributions to the experimental spectra was addressed
differently in all these investigations. We should also mention the
recent progress in radiative capture measurements~\cite{ham01:86}, where
the cross section has been measured at energies around 200 keV with an
accuracy of $\approx 15$ percent. Nevertheless, in all cases theoretical
models are needed to extrapolate the measured cross sections down to
solar energies. Theoretical studies of the low-energy behavior of the
astrophysical $S_{17}$-factor has been presented by many authors, see
e.g.~\cite{chr61:24,xu94:73,cso98:636,jen98:58,bar00:673,bay00:62,muk02:708}.

The structure of this paper is the following: Section~\ref{sec:theory}
contains a summary of the theoretical framework that will be used in the
calculations. In Sec.~\ref{sec:model} our analytical model WFs are
presented and discussed in quite some detail. Finally, in
Sec.~\ref{sec:emd} we discuss the model dependence of calculated EMD
energy spectra and propose two new methods to extract information from
EMD experiments.
\section{\label{sec:theory}Theoretical framework}
Our starting point for calculating electromagnetic cross sections will
be the $\uE\lambda$ strength function for a transition from a bound
state (total spin $J_0$) to a continuum state with energy $E$
%
\begin{multline}
  \dbde=\frac{1}{2J_0 + 1}
\\ \times
  \sum_{j} \int \ud \tau_j\left| \langle j||
 \mathcal{M}(\uE \lambda) || 0 \rangle \right|^2
  \delta\left(E_j-E\right),
\label{eq:strength}
\end{multline}
%
where $\ud \tau_j$ is the phase space element for final states,
$\mathcal{M} (\uE \lambda,\mu)$ is the electric multipole operator and
$| 0 \rangle$, $| j \rangle$ are the bound and continuum states in the
center of mass system.

We will consider loosely bound systems of two clusters $( c + x)$ and,
in particular, we will study transitions to the low-energy continuum in
which excitations are manifested as relative motion between the clusters
$E = \hbar^2 k^2 / 2 \mu_{cx}$, where $\mu_{cx}$ is the reduced mass of
the two-body system. Introducing the intercluster distance $r$, the
corresponding cluster $\uE \lambda$ operator (operating only on the
relative motion of clusters) is
\begin{equation}
  \mathcal{M}(\uE \lambda,\mu) = e Z(\lambda) r^\lambda Y_{\lambda \mu}
  (\hat{r}),
\label{eq:clusterel}
\end{equation}
where we have also introduced the effective multipole charge $Z(\lambda)
= \mu_{cx}^\lambda ( Z_x / m_x^\lambda + (-1)^\lambda Z_c /
m_c^\lambda)$.

The strength function is the key to study several reactions. For
example, the cross section for photo-dissociation $A(\gamma,x)c$ is
given by
\begin{equation}
  \sigma_\gamma^{\uE\lambda} (E) = \frac{(2\pi)^3 (\lambda +
  1)}{\lambda [ (2\lambda + 1)!!]^2} \left( \frac{E_\gamma}{\hbar c}
  \right)^{2\lambda -1} \dbde,
\label{eq:xsecgamma}
\end{equation}
where the photon energy $E_\gamma = E + E_0$ is larger than the binding
energy $E_0$. From this formula the inverse radiative capture reaction
$c(x,\gamma)A$ can be studied using detailed balance
\begin{equation}
  \sigma_\mathrm{rc}^{\uE\lambda} (E) = \left(
  \frac{E_\gamma}{\hbar c k} \right)^2 \frac{2 (2J_A +1)}{(2J_c + 1)
  (2J_x + 1)} \sigma_\gamma^{\uE\lambda} (E),
\label{eq:xsecrc}
\end{equation}
where $J_i$ is the spin of particle $i$. Note that the probability for
direct capture of charged particles is dramatically reduced at low
energies due to the Coulomb barrier in the $c + x$ channel. The cross
section is therefore usually factorized into the Gamow penetration
factor and the $S$-factor
\begin{equation}
  \sigma_\mathrm{rc} (E) = \frac{e^{-2\pi\eta(k)}}{E} S(E),
\label{eq:sfactordef}
\end{equation}
where $\eta(k) = Z_c Z_x e^2 \mu_{cx} / \hbar^2 k$ is the Sommerfeld
parameter. The dominant part of the energy behavior is carried by the
Gamow penetration factor while, e.g., nuclear structure information is
incorporated into the $S$-factor.

Finally, we will consider the process of EMD on a high-$Z$ target.
Using first-order perturbation theory, and the method of virtual
quanta~\cite{win79:319,ber85:442}, the energy spectrum can be written
as a sum over multipole $(\pi \lambda)$ photo-dissociation cross
sections multiplied by the corresponding spectra of virtual photons
$n_{\pi \lambda} (E_\gamma)$
%
\begin{multline}
  \frac{\ud \sigma_\uEMD}{\ud E} = \sum_{\uE\lambda} \frac{n_{\uE
  \lambda} (E_\gamma)}{E_\gamma} \sigma_\gamma^{\uE\lambda} (E)
\\
  + \sum_{\mathrm{M}\lambda} \frac{n_{\mathrm{M} \lambda}
  (E_\gamma)}{E_\gamma} \sigma_\gamma^{\mathrm{M}\lambda} (E).
\label{eq:xsecemd}
\end{multline}
%
Note that, except in the vicinity of corresponding resonances,
M$\lambda$ transitions are usually strongly
suppressed~\cite{ber85:442}. Therefore, we will not study them in this
work.
\section{\label{sec:model}Analytical model}
\subsection{Model wave functions}
A straightforward calculation of the electric multipole matrix element
for a direct transition between a loosely bound state and a non-resonant
continuum state, shows that the radial integrand rises to a maximum
value at a radius which is, in most cases, many times the nuclear
radius. Thus, these processes will mainly probe the surface structure of
the nucleus. Furthermore, ``loosely bound'' implies that the nucleus
will exhibit a large degree of clusterization and that the relative
motion WF between the core and the valence nucleon will have an extended
tail.

The final, continuum state will contain both Coulomb and nuclear
distortions. For low continuum energies, and when the binding energy of
the initial state is small, the nuclear distortions can be neglected in
a first approximation. Therefore, we will only consider a pure Coulomb
continuum in our analytical model, i.e., all nuclear phase shifts will
be put equal to zero. The effects of nuclear distortions in \bei\ EMD
will be considered in Sec.~\ref{sec:emdmodeldep}. Thus, a continuum
state, with relative orbital momentum $l$ between the clusters, will be
described by a normalized, regular Coulomb function
\begin{equation}
  \phi_{l} (k,r) = \sqrt{\frac{2}{\pi}} \frac{1}{k} i^{l} e^{i
  \sigma_{l}} F_{l} (k,r),
\label{eq:coulombwavefull}
\end{equation}
where
%
\begin{multline}
  F_{l} (k,r) = C_{l}(k) e^{ikr} (kr)^{l + 1}
\\ \times 
  {_1}F_1(l+1+i\eta (k) ; 2l + 2 ; -2ikr),
\label{eq:coulombwave}
\end{multline}
%
and $\sigma_{l}$ is the Coulomb phase, $\eta (k)$ is the Sommerfeld
parameter, ${_1}F_1(a;b;z)$ is the confluent hypergeometric
function~\cite{abr72}, and
\begin{equation}
C_{l} (k) = 2^{l} e^{-\pi \eta (k) / 2} |\Gamma(l + 1
+ i\eta (k))| / (2l + 1)!.
\label{eq:c}
\end{equation}

The reduced matrix element introduced in the definition of the strength
function, Eq.~\eqref{eq:strength}, contains a radial integral. With our
approximation for the continuum state this integral takes the form
%
\begin{multline}
  I_{l} (k) = \int_0^\infty \ud r e^{-i k r} r^{l + 1} 
\\ \times 
{_1}F_1 (l
  + 1 - i \eta (k) ; 2l + 2 ; 2 i k r) r^\lambda \phi (0,r).
\label{eq:radint}
\end{multline}
%
Here, $\phi (0,r)$ is the two-body, relative motion WF describing the
initial, bound state. At large $r$, with relative orbital momentum $l_0$
between the clusters, this radial function should be proportional to the
Whittaker function $W_{-\eta_0,l_0+1/2} (2 \kappa_0 r)$, see
e.g.~Ref.~\cite{abr72}, where $\eta_0 = Z_c Z_x e^2 \mu_{cx} / \hbar^2
\kappa_0$ and $E_0 = \hbar^2 \kappa_0^2 / 2 \mu_{cx}$ is the binding
energy.

In most studies on loosely bound systems, the Whittaker function has
been used to describe the bound state for all $r$. However, the
Whittaker function behaves as $r^{-l_0}$ in the limit $r \to 0$, and
therefore this approximation is only motivated if the transition matrix
element is dominated by contributions from very large $r$. This is the
case for reactions at very small energies; while for real experimental
energies ($E \gtrsim 100$~keV), the WF of the bound state should be
constructed in a more realistic way. Our idea is therefore to introduce
a model function that describes the bound state $(c + x)$ WF accurately
for all distances. This can be achieved by considering the behavior at
small and large $r$. We have already pointed out that the WF should be
described by a Whittaker function at large $r$. Furthermore, the
expected $r \rightarrow 0$ behavior for a two-body system consisting of
point-like particles is $r^{l_0 + 1}$. Both asymptotics are fulfilled
using the following model function
%
\begin{multline}
  \phi_{\tilde{\gamma}}^\mathrm{``exact\textrm{''}} (0,r) =
  \frac{1}{\sqrt{N_{\tilde{\gamma}}}}
  W_{-\eta_0,l_0+1/2} (2 \kappa_0 r) 
\\ \times
\left(1 - e^{-\kappa_1 r}
  \right)^{2l_0 + 1},
\label{eq:exactmodelwf}
\end{multline}
%
where $N_{\tilde{\gamma}}$ is the normalization constant and
$\tilde{\gamma}$ denotes the parameters $\left\{ \kappa_0, \kappa_1,
\eta_0 \right\}$. The parameters $\kappa_0$ and $\eta_0$ are defined by
the binding energy, charges, and masses, while $\kappa_1$ can be fitted
to give the correct distance between particles $c$ and $x$ (or the
correct size of the system). Using this WF, and solving the
integral~\eqref{eq:radint} numerically, it is possible to get very good
estimates for the electromagnetic reaction cross sections. We should
also mention that in the limit $k \to 0$ we are actually able to solve
the radial integral~\eqref{eq:radint} analytically, except for the
normalization constant $N_{\tilde{\gamma}}$.

However, we are searching for a completely analytical model which will
also enable us to incorporate many-body effects. Our model function has
to be modified accordingly.  First, we note that the asymptotic form of
the Whittaker function as $r \rightarrow \infty$ is
%
\begin{multline}
  W_{-\eta_0,l_0+1/2} (2 \kappa_0 r) \sim \frac{e^{-\kappa_0
  r}}{(2 \kappa_0 r)^{\eta_0}} 
\\ \times
  \left[ 1 - \frac{(\eta_0 - l_0) + (\eta_0^2 - l_0^2)}{2 \kappa_0 r} +
  \mathcal{O} \left( \frac{1}{r^2} \right) \right].
\label{eq:asympwhit}
\end{multline}
%
Secondly, for two-body systems in which the clusters have an internal
structure, the centrifugal barrier is effectively larger and the WF
should behave as $r^n$ (where $n > l_0 + 1$) as $r \rightarrow 0$.

Motivated by this, we put forward the following model function
\begin{equation}
  \phi_{\gamma} (0,r) = \frac{1}{\sqrt{N_\gamma}}
  \frac{e^{-\kappa_0 r}}{r^{\eta_0'}} \left(1 - e^{-\kappa_1 r}
  \right)^p,
\label{eq:modelwf}
\end{equation}
with norm
\begin{equation}
  N_\gamma = \sum_{m=0}^{2p} \binom{2p}{m} (-1)^m
  (2\kappa_0 + m\kappa_1)^{2\eta_0' - 1} \Gamma(1 - 2\eta_0'),
\label{eq:modelwfnorm}
\end{equation}
where $\gamma$ denotes the parameters $\left\{ \kappa_0, \kappa_1,
\eta_0', p \right\}$, and $p$ is an integer fulfilling $p > \eta_0' +
1$. The parameter $\kappa_0$ is defined by the binding energy and
effective mass. By putting $\eta_0' = \eta_0$ we would ensure to
reproduce the tail of the WF at very large $r$. However, the difference
between an exact Whittaker function and its asymptotic behavior (first
term of Eq.~\eqref{eq:asympwhit}) remains important for $r \lesssim
100$~fm. Therefore, $\eta_0'$ and $\kappa_1$ are used as free parameters
in a fit to the ``exact'' WF~\eqref{eq:exactmodelwf} in the interval of
interest. In this way $\eta_0'$ will be an \emph{effective} Sommerfeld
parameter while $\kappa_1$ will still mainly be connected with the
size. Note that if $\eta_0 > l_0$, then the second term in
Eq.~\eqref{eq:asympwhit} will be negative and consequently we will find
that $\eta_0' < \eta_0$. Finally, the integer $p$ is fixed by the small
$r$ behavior. For a pure two-body system we will use $p = <\eta_0' + l_0
+ 1>$ (where $<x>$ is the closest integer to $x$), while we can take
many-body effects into account by putting $p = <\eta_0' + n>$.

With this model WF it is possible to solve the integral~\eqref{eq:radint}
exactly
\begin{multline}
  I_{l,\gamma} (k) = \frac{1}{\sqrt{N_\gamma}} \sum_{m=0}^{p}
  \binom{p}{m} (-1)^m 
\\ \times
  (m \kappa_1 + \kappa_0 + i k)^{-(l + 2 + \lambda
  - \eta_0')} \Gamma (l + 2 + \lambda - \eta_0')
\\ \times 
  {_2}F_1 \Big( l + 2 + \lambda - \eta_0' ; l + 1 - i \eta (k) ; 
\\
  2 l + 2 ; \frac{2 i k}{m\kappa_1 + \kappa_0 + i k} \Big).
\label{eq:anaint}
\end{multline}
Many-body nuclear structure can further be taken into account by
considering the possibility that the bound state WF contains several
different two-body components
\begin{equation}
  \phi(0,r) = \sum_i a_i \phi_{\gamma_i} (0,r),
\label{eq:comp}
\end{equation}
which can be seen as two-body projections of the many-body WF. Note that
pure many-body components will not contribute to two-body break\-up and,
as a result, we will have $\sum_i a_i^2 < 1$. Note also that the
threshold for two-body breakup will be higher for components where one
(or both) of the clusters is excited. Therefore, we define the continuum
strength function separately for each component. Finally, we arrive at
an analytical formula for the strength function of component $i$
%
\begin{multline}
  \left. \dbde \right|_i = \frac{e^2 Z^2(\lambda) \mu_{cx}}{\hbar^2}
  \frac{2\lambda + 1}{2\pi^2} 
\\ \times
\sum_{l} a_i^2 k^{2 l + 1} C_{l}^2
  (k) \langle l_0 0 \lambda 0 | l 0 \rangle^2 | I_{l,\gamma_i} (k)
  |^2.
\label{eq:anastrength}
\end{multline}
%
\subsection{One-neutron halo systems}
The special case where $Z_x = 0$, i.e., a one-neutron halo system will
lead to several simplifications. First of all we will have $\eta_0 = 0$
and the Whittaker function in Eq.~\eqref{eq:exactmodelwf} will transform
into a modified, spherical Bessel function
\begin{equation}
  W_{0,l_0+1/2} (2 \kappa_0 r) = \sqrt{\frac{2 \kappa_0 r}{\pi}} K_{l_0
  + 1/2} (\kappa_0 r).
\label{eq:exactmodelwf1n}
\end{equation}
Furthermore, the continuum solution will reduce from a Coulomb
function~\eqref{eq:coulombwavefull} to the corresponding component of a
plane wave
\begin{equation}
  \phi_{l}^{(1n)} (k,r) = \sqrt{\frac{2}{\pi}} i^{l} r j_{l} (kr),
\label{eq:planewave}
\end{equation}
where $j_l(x)$ is a spherical Bessel function. In this case, the
integral~\eqref{eq:radint} can be solved exactly. Consider, e.g., a
node-less $s$ state, for which our model WF would read
\begin{equation}
  \phi_{\tilde{\gamma}}^{(1n,1s)} (0,r) =
  \frac{1}{\sqrt{N_{\tilde{\gamma}}^{(1n,s)}}}
  e^{-\kappa_0 r} \left( 1 - e^{-\kappa_1 r} \right), 
\label{eq:1nl0modelwf}
\end{equation}
with norm
\begin{equation}
  N_{\tilde{\gamma}}^{(1n,1s)} = \frac{\kappa_1^2}{2\kappa_0
  (\kappa_1+2\kappa_0)(\kappa_1+\kappa_0)}.
\label{eq:1nl0modelwfnorm}
\end{equation}
The radial integral~\eqref{eq:radint} for this special case will be
given by
%
\begin{multline}
  I_{l,\tilde{\gamma}}^{(1n,1s)} (k) =
  \frac{1}{\sqrt{N_{\tilde{\gamma}}^{(1n,1s)}}}
  \frac{(2l)!!(l+1+\lambda)!}{2^l l!} 
\\ \times
  \sum_{m=0}^{1} (-1)^m (m \kappa_1
  + \kappa_0)^{-(l + 2 + \lambda)} 
\\ \times
  {_2}F_1 \left( \frac{l + 2
  + \lambda}{2} ; \frac{l + 3 + \lambda}{2} ; \frac{3}{2} + l ;
  \frac{-k^2}{(m\kappa_1 + \kappa_0)^2} \right).
\label{eq:anaint1nl0}
\end{multline}
%
\subsection{Application to $\mathbf{\bei}$%
\label{sec:b8wf}}
We will now apply our model to the \bei\ nucleus. The low-lying \bei\
continuum can, with relatively good precision, be approximated as a pure
Coulomb one. At least there are no negative parity states at low
excitation energies~\cite{ajz88:490,gol98:67,rog01:64} and the
electromagnetic processes are, in all cases we are considering,
dominated by $\uE 1$ transitions. However, we will also show that the
influence of a broad negative parity state at high excitation
energy~\cite{gol98:67,rog01:64} is not negligible at intermediate
($\gtrsim 0.4$~MeV) continuum energies.

\begin{figure}[hbt]
  \includegraphics[width=80mm]{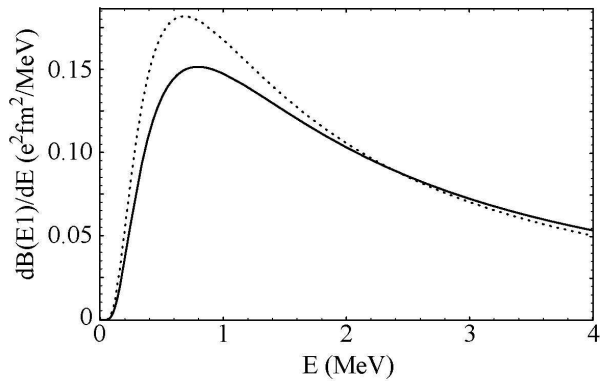}
  \caption{The E1 strength function of \bei. Although the total strengths
  are the same in these examples, the shapes are very different when
  treating the \bei\ nucleus as a two-body (dotted line) or a three-body
  (solid line) system. The difference is mainly due to the stronger
  centrifugal barrier in the three-body case (see, e.g.,
  Ref.~\cite{del60:20}) which forces the WF to be 
  narrower in coordinate space and thus wider in momentum/energy
  space. The parameters of the model WFs can be found in
  Table~\ref{tab:parameters}.%
  \label{fig:e1strength}}
\end{figure}

In a first approximation the bound state of \bei\ can be treated as a
pure two-body ($\bese + p$) system with binding energy $E_0 = 137$~keV
and relative orbital momentum $l_0 = 1$. The single free parameter
$\kappa_1$ in our ``exact'' model function~\eqref{eq:exactmodelwf} was
then fitted to an rms intercluster distance of $r_\urms = 4.57$~fm
(extracted from Ref.~\cite{gri99:60}). In order to get analytical
results, we then introduced the model function~\eqref{eq:modelwf}. We
put $p = 3$ and fixed $\kappa_0$ from the binding energy, while the
remaining parameters $\kappa_1$ and $\eta_0'$ were fitted to the
behavior of the ``exact'' model WF, see Table~\ref{tab:parameters}. The
resulting E1 strength function is shown as a dotted line in
Fig.~\ref{fig:e1strength}. This analytical approximation agrees very
well with numerical results obtained keeping the ``exact'' model WF. The
error is less than 2\% in the region of interest.
%
\begin{table}[hbt]
  \caption{Parameters of our model WFs used to describe the \bei\ ground
  state. Both models give the same intercluster distance, $r_\urms =
  4.57$~fm. With $E_0 = 137$~keV we get $\kappa_0 = 0.076$~fm$^{-1}$ and
  $\eta_0 = 1.595$. The excited core component (last row) has $E_0 =
  566$~keV giving $\kappa_0 = 0.154$~fm$^{-1}$ and $\eta_0 = 0.786$. The
  relative orbital momentum for all components are $l_0 =1$ while $I$ is
  the channel spin and $a^2$ is the spectroscopic factor. Note that
  there is no dependence on the particular value of the channel spin in
  the two-body case.%
  \label{tab:parameters}}
\begin{ruledtabular}
\begin{tabular}{l | l c c c c c}
      Model WF & configuration & $I$ & $a^2$ & $p$ & $\kappa_1$~(fm$^{-1}$) &
      $\eta_0' / \eta_0$ \\
	\hline
      	two-body & $[ \bese(3/2^-) \otimes p ]$ & 2 &
    	1.00 & 3 & 0.601 & 0.79 \\
	\hline
	& $[ \bese(3/2^-) \otimes p ]$ & 2 & 0.65 & 5 &
    	0.702 & 0.87 \\
	& $[ \bese(3/2^-) \otimes p ]$ & 1 & 0.13 & 5 &
    	0.765 & 0.86 \\
	\raisebox{6ex}[0pt]{three-body} & $[ \bese(1/2^-)
    	\otimes p ]$ & 1 & 0.16 & 5 & 0.753 & 1.43 \\
  \end{tabular}
\end{ruledtabular}
\end{table}

However, concerning the structure of the \bei\ ground state one should
keep in mind that the \bese\ core is in itself a weakly bound system
with an excited $1/2^-$ state at 429~keV. The common treatment of \bei\
as a pure two-body system is therefore questionable. We want to
investigate what effect the many-body structure of \bei\ might have on
the strength function. For this purpose we utilize a recent three-body
($\alpha + {}^3\mathrm{He} + p$) calculation~\cite{gri99:60} where it
was shown that, after projection onto the two-body channels, there are
three main components (adding up to 94\% of the total WF, see
Table~\ref{tab:parameters}) and that the rest are pure three-body
channels. For each of the numerical two-body overlap functions we fit
our parameters $\kappa_1$ and $\eta_0'$. The binding energy, $E_0 =
137$~keV, determines $\kappa_0$ for the two first components and $E_0 =
566$~keV for the third, \bese\ excited state, component. The best fit to
the small $r$ behavior is obtained with $p = 5$ which reflects the
effectively larger centrifugal barrier in the three-body case. This
centrifugal barrier will push the WF away from $r=0$ and will, thus,
force it to become more narrow than the corresponding two-body WF.
\subsection{Studies of the corresponding potential}
\begin{figure}[hpbt]
  \includegraphics[width=80mm]{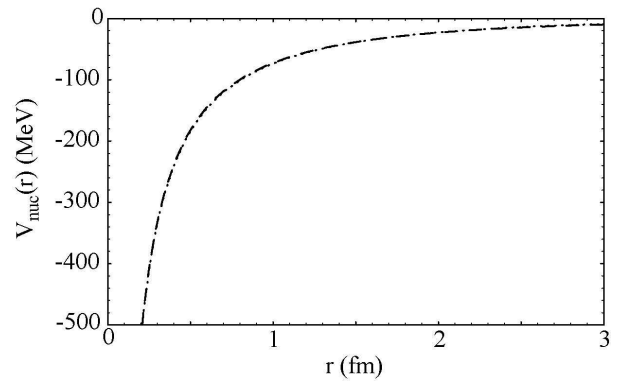}
  \caption{The nuclear potential (dotted line) corresponding to our
  \bei\ ``exact'' WF. This potential is very well described by a
  Yukawa potential $V_\mathrm{nuc}(r) = -114 e^{-0.457 r}/r$~(MeV)
  (dashed line).%
  \label{fig:pot}}
\end{figure}

Using the two-body WF~\eqref{eq:exactmodelwf}, which describes correctly
the binding energy and the geometry, we are able to restore the
corresponding two-body potential. Besides centrifugal barrier and
Coulomb interaction, the potential also contains an attractive part
which we find can be approximated with a high accuracy by one or two
Yukawa-type potentials. Note that such potentials are widely used in
few-body nuclear physics. In Fig.~\ref{fig:pot} the nuclear part of the
potential which corresponds to our \bei\ ``exact'' two-body WF is
plotted. As is shown in the figure, the potential can be very well
described by a Yukawa-type potential. In this connection we would also
like to point out that for the special case where $l_0=0$ and $Z_x = 0$,
the WF can be described by~\eqref{eq:1nl0modelwf} and the corresponding
potential reduces to a Hulth\'en potential which has exact solutions.
%
\section{\label{sec:emd}Electromagnetic dissociation}
The EMD of loosely bound nuclei, impinging on a high-$Z$ target, has
been used in nuclear physics for many years both in order to investigate
nuclear structure and as an indirect method to extract information on
radiative capture reactions. Unfortunately, as for all reaction
experiments, a lot of information is contained in the experimental
results and it is a hard task to disentangle the desired part. The
transition matrix elements represents the probability for an initial
state wave function to end up in a specific final state after being
filtered through the reaction mechanism. Naturally, both the structure
of the initial as well as the final state are important for this
quantity. In addition, further complications arise if nuclear induced
breakup contributes to the measured cross section, and/or if the
interaction time is long enough for higher-order transitions to become
important. Therefore, in order to minimize interference from
higher-order dynamical effects and from nuclear interactions, we will be
interested in high-energy experiments in which events characterized by
large impact parameters have been selected.

In this section we will demonstrate that the analysis of EMD energy
spectra from loosely bound nuclei is highly model dependent. We will
then discuss the important issue of how to separate the contributions
from specific multipoles; a problem which is of great significance
when extracting information on the inverse, radiative capture
reaction.
\subsection{Model dependence of energy spectrum analysis
\label{sec:emdmodeldep}}
Applying our analytical model to study EMD using first-order
perturbation theory enables us to investigate the sensitivity of the
energy spectrum to some properties of the initial bound state. At small
relative energies the electromagnetic processes are highly peripheral,
which means that the interaction mainly probes the external part of the
bound state WF. As a consequence, the amplitude of the EMD cross section
should depend crucially on the size of the nuclear system. A larger size
implies a lower Coulomb barrier, which results in a larger tunneling
probability, and consequently a larger cross section. However, the radii
of nuclei far from stability are usually extracted from interaction
cross section measurements, and this procedure has unfortunately proven
to be highly model dependent. A Glauber-type analysis, assuming a
uniform density distribution, results in a smaller radius as compared to
an analysis in which the few-body structure is taken into account
explicitly~\cite{alk96:76}. Furthermore, the relevant parameter for
two-body breakup is in reality the intercluster distance rather than the
total matter radius, and the relation between these two quantities is
also model dependent. In a pure two-body model one often assumes that
the size of the core is equal to the size of the corresponding free
nucleus. In contrast, taking many-body structure into account will
result in polarization effects. For example, it was found in
Ref.~\cite{gri98:57} that the average distance between
${^3}\mathrm{He}-\alpha$ is approximately 10\% smaller inside \bei\
(studied in a ${^3}\mathrm{He}+\alpha+p$ picture) than in a free \bese\
nucleus (${^3}\mathrm{He}+\alpha$ picture).

\begin{figure}[hpbt]
  \includegraphics[width=80mm]{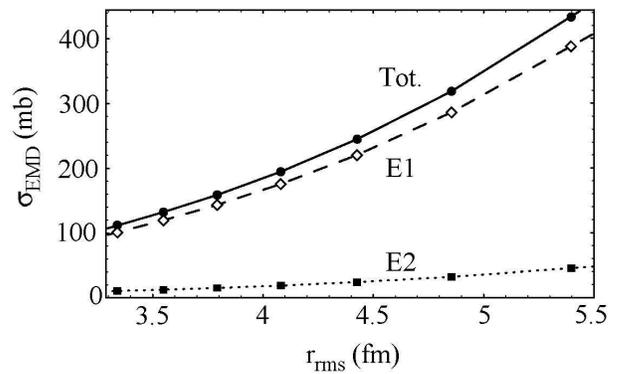}
  \caption{The total EMD cross section as a function of intercluster
  distance. The parameter $\kappa_1$ of our two-body model WF was varied
  in order to change the rms distance between the \bese\ core and the
  valence proton. A beam energy of 82.7~MeV/nucleon and a minimum impact
  parameter of 30~fm was used.%
  \label{fig:stot}}
\end{figure}

We have investigated the sensitivity of the EMD cross section to the rms
intercluster distance by performing model calculations with a \bei-like
system. A pure two-body system with $l_0 = 1$ relative motion (having
unity spectroscopic factor) was assumed and a binding energy of
0.137~MeV was used. The effective Sommerfeld parameter was fitted to the
asymptotic behavior of~\eqref{eq:exactmodelwf}, i.e., to the Whittaker
tail. The size of our model WF~\eqref{eq:modelwf} was then a function of
the remaining free parameter $\kappa_1$. By varying this parameter we
could investigate the sensitivity to the size and from
Fig.~\ref{fig:stot} it is clear how strong it is. Just for illustration:
a 15\% uncertainty in the rms intercluster distance (which, in our
model, would correspond to 5\% uncertainty in the \bei\ matter radius)
results in $\sim 50$\% ambiguity of the calculated total cross section.

Let us now consider the difference between a two-body and a three-body
approach. As was mentioned in Sec.~\ref{sec:b8wf}, the effectively
larger centrifugal barrier in a three-body system will push the relative
motion WF away from $r=0$ and consequently force it to become more
narrow than the corresponding two-body WF for a given radius. We
therefore expect the distribution in momentum/energy space to be
broader. This effect is clearly seen in Fig.~\ref{fig:e1strength} where
the \bei\ E1 strength functions, obtained using our
three-body~\footnote{Note that this is not strictly a three-body model,
but rather the two-body projection of a three-body WF. However, in the
following we will consistently refer to it as three-body results.} and
two-body analytical model WFs, are compared. This difference, seen in
the strength function, should be even more pronounced in the energy
spectrum since it will be magnified by the spectrum of virtual photons.

\begin{figure}[hbt]
  \includegraphics[width=80mm]{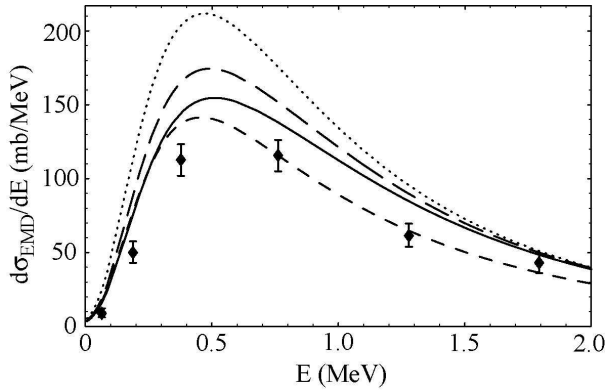}
  \caption{The \bei\ EMD energy spectrum obtained at 82.7~MeV/nucleon on
  Pb with \bei\ scattering angles $\theta_{\bei} \leq 1.77^\circ$. The
  data points are from~\cite{dav01:63}. The curves show our analytical
  two-body (dotted line) and three-body (solid line) results, and the
  numerical calculations: without FSI (long-dashed), and with FSI
  (short-dashed). All theoretical curves have been corrected for
  experimental resolution and acceptance.%
  \label{fig:dsde}}
\end{figure}

In Fig.~\ref{fig:dsde} we compare different calculations of \bei\ EMD on
Pb, including both E1 and E2 transitions, to the experimental data from
Davids~{et~al.}~\cite{dav01:63}. This experiment is very appealing since
the selection of scattering angles ($\theta_{\bei} \leq 1.77^\circ$,
which corresponds to a minimum impact parameter of $b_\mathrm{min} =
30$~fm) minimizes the contribution from nuclear scattering, and the
relatively high beam energy (82.7~MeV/nucleon) justifies the use of
first-order perturbation theory. Let us first compare our analytical
two-body (dotted line) and three-body (solid line) results, see also
Ref.~\cite{for02:549}. Concerning the shape of the energy spectrum we
have an excellent agreement between the experimental data and our
results obtained using the three-body model, while the pure two-body
calculation gives a too narrow peak. As to the absolute values, the
three-body energy spectrum is approximately 20\% above the experimental
data. However, the most important lesson from this comparison is that
for two different assumptions concerning the nuclear structure, but
keeping the rms intercluster distance fixed, we obtain very different
shapes of the calculated energy spectra. Thus, one can conclude that the
interpretation of energy spectra is highly model dependent. A final
remark in connection to this observation is that, in order to interpret
experimental data correctly, it is very important to fix the
spectroscopic factors of different two-body and many-body
components. Therefore, we want to stress the usefulness of experiments
where EMD is studied in complete kinematics. Examples of interesting
channels in the \bei\ case is $\bei \rightarrow \bese(1/2^-) + p +
\gamma$ and $\bei \rightarrow {^3}\mathrm{He} + \alpha + p$. Some progress
has already been made in this direction. Recently, \bese\ fragments and
$\gamma$-rays were measured in coincidence after breakup on a light
target by Cortina-Gil~{et~al.}~\cite{cor02:529} which resulted in a
clear observation of the excited core component of the WF.

We have also performed numerical calculations, based on first-order
perturbation theory, where the numerical \bei\ bound state WF
from~\cite{gri99:60} was used. In the first investigation we assumed a
pure Coulomb continuum and the results of this calculation is shown as a
long-dashed line in Fig.~\ref{fig:dsde}. We note that the obtained
energy spectrum compares rather well with our analytical three-body
model except for a 15\% difference in the peak height. Since our
analytical model has the correct asymptotic behavior for large $r$ and a
three-body behavior at small $r$, the main difference to the numerical
WF should be in the intermediate region, and this is exactly the region
which dominates the transition matrix elements for energies
corresponding to the peak of the energy spectrum. This fact explains the
observed discrepancy.

In our second numerical investigation we studied the influence of $\bese
+ p$ FSI. As previously mentioned, the low-energy continuum of \bei\ is
dominated by positive parity states~\cite{ajz88:490} which are only
relevant for E2 ($\sim$10\% of the total cross section) and M1
transitions. Note that the latter only plays a role in the vicinity of
the narrow $1^+$ resonance at 0.64~MeV above threshold, and is therefore
not included in our calculations. However, the possible existence of a
very broad negative parity state at high excitation energy can still
have a strong influence on the energy spectrum. Effects of such a state
was observed in a recent elastic proton scattering
experiment~\cite{rog01:64} from which the authors made a $2^-$
spin-parity assignment and, from an $R$-matrix analysis, they obtained a
best fit with the parameters $E = 3.5 \pm 0.5$~MeV and $\Gamma = 8 \pm
4$~MeV. We have included such a broad continuum structure by adding an
attractive potential in the $s$-wave channel. The effects of this are
clearly seen in Fig.~\ref{fig:dsde} (short-dashed line): the total cross
section is reduced and the continuum strength is redistributed towards
smaller energies. We conclude this comparison by stating that a broad
negative parity structure in the high energy continuum has a
non-negligible influence on the EMD energy spectrum for energies
$\gtrsim 0.4$~MeV and that the parameters of such a state are still to
be determined with greater accuracy.
\subsection{Extraction of E1 contribution using the low-$\bm{E}$
  energy spectrum}
As we have seen in the previous section the main model uncertainties in
the EMD analysis are: the asymptotic normalization constant which
depends on (i) the radius in combination with (ii) the spectroscopic
factors of different two-body components; (iii) the underlying many-body
structure, and finally; (iv) the FSI. Despite the difficulties connected
with measuring the cross section at small relative energies, we still
suggest to focus on the low-$E$ part of the energy spectrum. In this way
one can avoid uncertainties associated with FSI (unless there are
resonances very close to threshold) and with the many-body behavior of
the WF at small intercluster distances. The asymptotic normalization
constant in combination with the spectroscopic factor will enter as an
absolute normalization of the cross section. However, since this
normalization affects all multipole transitions equally it is possible
to calculate the ratio of two different multipoles with a very good
precision.

\begin{figure}[hpbt]
  \includegraphics[width=80mm]{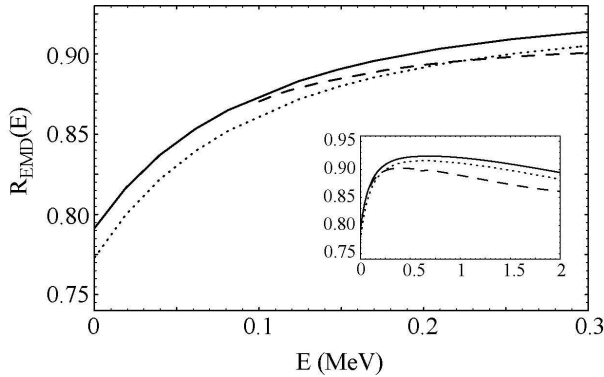}
  \caption{Fraction of the calculated energy spectrum (EMD 
  of 82.7~MeV/nucleon \bei\ on Pb with scattering angles $\theta_\bei
  \leq 1.77^\circ$) attributed to E1 transitions, see
  Eq.~\eqref{eq:dsderatiodef}. The curves show two-body 
  (dotted line) and three-body (solid line) analytical results while the
  dashed line shows results of numerical calculations including FSI. The
  inset shows the same curves for a broader energy range.%
  \label{fig:dsderatio}}
\end{figure}

We have investigated the ratio
\begin{equation}
  R_\uEMD (E) \equiv \frac{\ud \sigma_\uEMD (\uE 1)/\ud E} {\ud \sigma_\uEMD
  (\uE 1 + \uE 2)/\ud E},
\label{eq:dsderatiodef}
\end{equation}
and found that it is almost model independent at small relative
energies. To demonstrate this we will continue to use the EMD of \bei\
(on a Pb target at 82.7 MeV/nucleon with $b_\mathrm{min} = 30$~fm) as an
example. Reusing the different model calculations from
Sec.~\ref{sec:emdmodeldep} we can investigate the sensitivity of the
ratio to different model assumptions. Firstly, in
Fig.~\ref{fig:dsderatio} we can see that the difference between a
two-body and a three-body approach is less than 3\% in the region of
interest. This low sensitivity can be explained by the fact that the
low-$E$ part of the spectrum mainly probes the large $r$ asymptotics of
the radial WF. Shown in Fig.~\ref{fig:dsderatio} are also results from
the numerical calculations introduced in Sec.~\ref{sec:emdmodeldep}. It
is clearly seen that the influence from FSI is almost negligible at
small relative energies where the numerical results and our analytical
three-body model seem to converge. Unfortunately, the numerical accuracy
of our calculations becomes questionable at small energies and the
calculated ratio is therefore only plotted down to 0.1~MeV. In this
context we want to stress that in our analytical model we are able to
calculate the relevant transition matrix elements for all energies,
including the limit $E \to 0$. In contrast, the numerical approaches
will run into problems for small energies since the continuum WF will be
extremely small at relevant intercluster distances.

\begin{figure}[hpbt]
  \includegraphics[width=80mm]{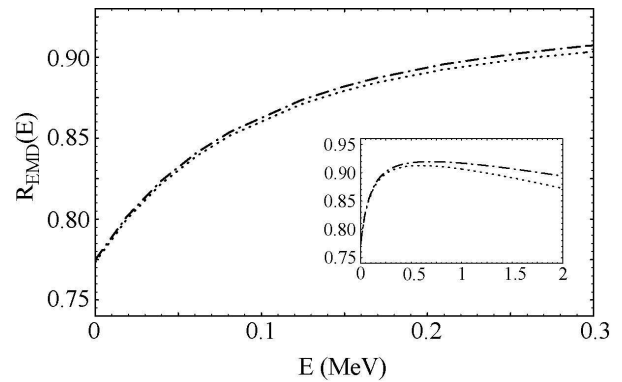}
  \caption{Fraction of the calculated energy spectrum (EMD 
  of 82.7~MeV/nucleon \bei\ on Pb with scattering angles $\theta_\bei
  \leq 1.77^\circ$) attributed to E1 transitions, see
  Eq.~\eqref{eq:dsderatiodef}. The curves show
  two-body results characterized by two different intercluster
  distances: $r_\mathrm{rms} = 3.3$~fm (dash-dotted) and $r_\mathrm{rms}
  = 5.4$~fm (dotted). The inset shows the same curves for a broader
  energy range.%
  \label{fig:dsderatiorad}}
\end{figure}

Furthermore, as can be seen in Fig.~\ref{fig:dsderatiorad}, the
sensitivity of the $R_\uEMD (E)$ ratio~\eqref{eq:dsderatiodef} to the
intercluster distance is also very small. This feature is also expected
since the intercluster distance determines the asymptotic normalization
of the WF which, in turn, cancels when the ratio is calculated. However,
as can be seen in the insets of Figs.~\ref{fig:dsderatio} and
\ref{fig:dsderatiorad}, the results obtained using the different models
diverge with increasing energy.

One final question is well-founded: Since the transition matrix elements
for very small energies depend mainly on the tail of the bound state WF,
is it still justified to use our model WF~\eqref{eq:modelwf} which has
merely an approximate description of the Whittaker tail? The result of
our numerical calculation (remember that the numerical bound state WF
has the correct asymptotics) presented in Fig.~\ref{fig:dsderatio}
indicates that it is justified, since it seems to converge with the
analytical model for small $E$. Furthermore, in the limit $E \to 0$ we
are actually able to solve the radial integral~\eqref{eq:radint} exactly
even for the ``exact'' WF~\eqref{eq:exactmodelwf} which has a Whittaker
tail. We find that $R_\uEMD (E \to 0)$ calculated with the model WF and
with the ``exact'' WF agree within 0.5\%. This result gives an
additional justification to the use of our model WF for calculating
transition matrix elements at small $E$.

\begin{figure}[hpbt]
  \includegraphics[width=80mm]{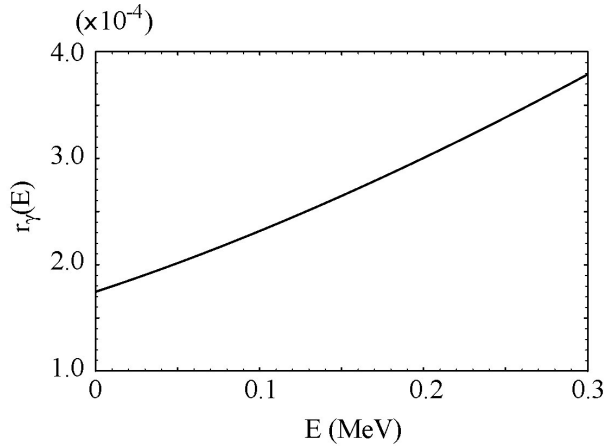}
  \caption{Ratio of photo-dissociation cross sections calculated
  within our analytical three-body model, see
  Eq.~\eqref{eq:sgphotoratiodef}.%
  \label{fig:sigphotoratio}}
\end{figure}

In summary we have found that the calculated $R_\uEMD (E)$ ratio of the
energy spectrum is almost model independent at small relative
energies. In first-order perturbation theory this ratio can be expressed
as
%
\begin{equation}
\begin{aligned}
  R_\uEMD (E) &\equiv
  \frac{\ud \sigma_\uEMD(\uE 1)/\ud E}{\ud \sigma_\uEMD(\uE 1+\uE 2)/\ud E}
\\
  &= 1 \Big/ \left( 1 + \frac{n_{\uE 2}(E_\gamma)}{n_{\uE 1}(E_\gamma)}
  \frac{\sigma_\gamma^{\uE 2}(E)}{\sigma_\gamma^{\uE
  1}(E)} \right) \\
  &= 1 \Big/ \left( 1 + \frac{n_{\uE 2}(E_\gamma)}{n_{\uE 1}(E_\gamma)}
  r_\gamma(E) \right),
\end{aligned}
\label{eq:dsderatio}
\end{equation}
where we have introduced the E2/E1 ratio of photo-dissociation cross
sections
\begin{equation}
  r_\gamma(E) \equiv \frac{\sigma_\gamma^{\uE 2}(E)}{\sigma_\gamma^{\uE
  1}(E)}.
\label{eq:sgphotoratiodef}
\end{equation}
Naturally, the ratio $R_\uEMD (E)$ will depend on the experimental
conditions such as beam energy and minimum impact parameter. However,
this dependence enters only in the spectra of virtual photons which are
easily calculated for any experimental conditions $\left\{
b_\mathrm{min}, E_\mathrm{beam} \right\}$ (see, e.g., Eq.~(4.7)
of~\cite{ber85:442}). In contrast, the ratio of \bei\ photo-dissociation
cross sections $r_\gamma(E)$~\eqref{eq:sgphotoratiodef} does not depend
on the experimental conditions.  Therefore, we provide this ratio,
calculated within our analytical three-body model, in
Fig.~\ref{fig:sigphotoratio}. The curve shown in
Fig.~\ref{fig:sigphotoratio} can approximately be described by the
formula
\begin{equation}
  r_\gamma(E) = \frac{\sigma_\gamma^{\uE 2}(E)}{\sigma_\gamma^{\uE
  1}(E)} = \left(-1.7290 + 3.4663 e^{1.5526 E} \right) \cdot 10^{-4},
\label{eq:approxsigphotoratio}
\end{equation}
which describes the calculated curve with an accuracy of $< 0.5$\% in
the region $0 < E < 0.3$~MeV. From this formula it is easy to obtain
$R_\uEMD (E)$ using Eq.~\eqref{eq:dsderatio}
%
\subsection{Extraction of E1 contribution using two different experimental
  conditions} 
One of the beauties with the method of virtual photons is the separation
of reaction kinematics and nuclear excitation dynamics into the spectrum
of virtual photons and the photo-absorption cross section,
respectively. This separation can be used as an alternative method to
extract the E1 contribution from the measured cross section. The objective
is to use the fact that the cross section will depend on beam energy
and minimum impact parameter only through the spectra of virtual
photons. First, let us introduce the notation
\begin{equation}
  N_{\uE\lambda} (E_\gamma) = \frac{n_{\uE\lambda}(E_\gamma,
  b_\mathrm{min}, E_\mathrm{beam})} {E_\gamma},
\label{eq:virtphoti}
\end{equation}
where we have indicated that the spectra of virtual photons are
functions of the beam energy and minimum impact parameter. In reality
this dependence enters in the adiabaticity parameter
\begin{equation}
  \xi = \frac{\omega b_\mathrm{min}}{\gamma v}.
\label{eq:adiabaticitymin}
\end{equation}
Assuming that the total cross section is dominated by E1 and E2
transitions we find that the energy spectrum for given experimental
conditions 
$\{ b_\mathrm{min}^{(i)}, E_\mathrm{beam}^{(i)} \}$ is given by 
\begin{equation}
  \frac{\ud \sigma_\uEMD^{(i)}}{\ud E} = N_{\uE 1}^{(i)} (E_\gamma)
  \sigma_\gamma^{\uE 1} (E) + N_{\uE 2}^{(i)} (E_\gamma)
  \sigma_\gamma^{\uE 2} (E),
\label{eq:xsecemdi}
\end{equation}
where we would like to remind the reader of the relation $E = E_\gamma -
E_0$. Now we can use the fact that the virtual photon spectra depends
differently on $\xi$ for different multipoles to extract the
contribution from one of the multipoles.  Let us assume that we have two
sets of experimental data from the same experimental setup; the only
difference being the beam energy and/or the selected scattering angles
(minimum impact parameter). The E1 contribution to one of the
measurements can then be obtained with the formula
%
\begin{multline}
  \frac{\ud \sigma_\uEMD^{(1)}(\uE 1)}{\ud E} = \left( N_{\uE
  2}^{(2)} (E_\gamma) \frac{\ud \sigma_\uEMD^{(1)}}{\ud E} - 
  N_{\uE 2}^{(1)} (E_\gamma) \frac{\ud \sigma_\uEMD^{(2)}}{\ud E}
  \right)\\
  \times \frac{N_{\uE 1}^{(1)}
  (E_\gamma)} { N_{\uE 1}^{(1)} (E_\gamma) N_{\uE 2}^{(2)} (E_\gamma) -
  N_{\uE 1}^{(2)} (E_\gamma) N_{\uE 2}^{(1)} (E_\gamma)}.
\label{eq:xsecemd1}
\end{multline}
The advantage of this method is that information can be obtained
directly from experimental data. However, it should be emphasized that
the formula is only valid under the assumptions of first-order
perturbation theory and straight-line trajectories. Thus, this method
can only be used for EMD at relatively large beam energies where events
characterized by large impact parameters have to be
selected. Furthermore, the experimental conditions must be chosen so
that the difference $\ud \sigma_\uEMD^{(1)}/\ud E - \ud
\sigma_\uEMD^{(2)}/\ud E$ is observable and larger than the experimental
uncertainty.
\section{\label{sec:conc}Conclusion}
In this paper we have studied electromagnetic processes involving
loosely bound nuclei. To this aim we have developed an analytical model
which is based on the use of radial model functions that give a
realistic description of two-body WFs (or the two-body projections of
many-body WFs) for all radii, see also~\cite{for02:549}. We have used
this model to study EMD of \bei, but have also indicated how it can be
applied to other reactions and other nuclei. For example, it should
provide an important tool to investigate the low-energy behavior of the
astrophysical $S$-factor.

We have also presented numerical calculations based on the three-body
model of \bei, developed in Ref.~\cite{gri99:60}, and on recent
experimental information on a broad negative parity state in the \bei\
continuum. Combining the results of our analytical model, and of these
numerical calculations, has allowed us to investigate the sensitivity of
calculated EMD energy spectra to different model assumptions. We
concluded that the magnitude of the cross section depends strongly on
the intercluster distance and on the spectroscopic factors of different
many-body channels; while the shape of the energy spectrum is very
sensitive to the few-body structure. Finally, we found that a broad
negative parity state at high excitation energy will influence the
energy spectrum for energies $\gtrsim 0.4$~MeV, and will lead to a
reduction of the cross section.

However, the main purpose of this paper has been to investigate the
problem of how to extract the E1 contribution from a measured EMD energy
spectrum. This question is of great significance for the gathering of
information on astrophysically interesting radiative capture reactions
from EMD experiments (note the relation between the E1 component of the
EMD energy spectrum and the astrophysical $S$-factor via
Eqs.~(\ref{eq:xsecrc}--\ref{eq:xsecemd})). The main method so far has
been to study the asymmetries in angular or momentum
distributions. However, this asymmetry (which is due to E1-E2
interference) depends strongly on details of the FSI which, in turn, are
often relatively unknown. Furthermore, the E1-E2 interference terms do
not themselves contribute to the integrated cross sections to which the
$S$-factor is related. Instead, we have proposed two novel, and less
model dependent, approaches to extract the E1 contribution from a
measured EMD energy spectrum: (i) Firstly, we demonstrated that the
ratio of EMD cross sections $\sigma_\uEMD (\uE 1)/\sigma_\uEMD (\uE
1+\uE 2)$ is almost model independent at small relative energies. We
also provided an analytical formula to calculate this ratio for any
experimental conditions. (ii) Secondly, we demonstrated how two sets of
experimental data, obtained with different $E_\mathrm{beam}$ and/or
$b_\mathrm{min}$, can be used to extract the E1 component. This method
relies on the fact that the strengths of different multipole components
depend on the beam energy and minimum impact parameter, and in
first-order perturbation theory this dependence enters only in the
virtual photon spectra.

Since the proposed two methods are not directly connected to each other,
they can be used independently and the results can be compared to each
other. However, both methods, but in particular the first one, require
that the energy spectrum is measured down to very small relative
energies (100--300~keV), which will probably prove to be a difficult
challenge.
\begin{acknowledgments}
N.~B.~S.~is grateful for support from the Royal Swedish Academy of
Science. The support from RFBR Grants No 00--15--96590, 02--02--16174
are also acknowledged.
\end{acknowledgments}

\end{document}